\documentclass[runningheads]{llncs}
\usepackage{graphicx}
\usepackage{amsmath}
\usepackage{amssymb}
\usepackage{array}
\usepackage[misc]{ifsym}
\begin{document}
\title{Invertible Sharpening Network for MRI Reconstruction Enhancement}
\titlerunning{Invertible Sharpening Network for MRI Reconstruction Enhancement}
%
\author{Siyuan Dong\thanks{This work was done during the internship at United Imaging Intelligence}\inst{1} \and Eric Z. Chen\inst{2} \and Lin Zhao\inst{3} \and Xiao Chen\inst{2} \and Yikang Liu\inst{2} \and Terrence Chen\inst{2} \and Shanhui Sun\inst{2}\textsuperscript{(\Letter)}}

%
\authorrunning{S. Dong et al.}
%
\institute{Electrical Engineering, Yale University, New Haven, CT, USA
\email{s.dong@yale.edu}\and
United Imaging Intelligence, Cambridge, MA, USA
\email{shanhui.sun@uii-ai.com}\and
Computer Science, University of Georgia, Athens, GA, USA}
\maketitle              
\begin{abstract}
High-quality MRI reconstruction plays a critical role in clinical applications. Deep learning-based methods have achieved promising results on MRI reconstruction. However, most state-of-the-art methods were designed to optimize the evaluation metrics commonly used for natural images, such as PSNR and SSIM, whereas the visual quality is not primarily pursued. Compared to the fully-sampled images, the reconstructed images are often blurry, where high-frequency features might not be sharp enough for confident clinical diagnosis. To this end, we propose an invertible sharpening network (InvSharpNet) to improve the visual quality of MRI reconstructions. During training, unlike the traditional methods that learn to map the input data to the ground truth, InvSharpNet adapts a backward training strategy that learns a blurring transform from the ground truth (fully-sampled image) to the input data (blurry reconstruction). During inference, the learned blurring transform can be inverted to a sharpening transform leveraging the network's invertibility. The experiments on various MRI datasets demonstrate that InvSharpNet can improve reconstruction sharpness with few artifacts. The results were also evaluated by radiologists, indicating better visual quality and diagnostic confidence of our proposed method. 

\keywords{MRI Recon  \and Sharpness Enhancement \and Invertible Networks}
\end{abstract}
\section{Introduction}
Due to the hardware limitations, the Magnetic Resonance Imaging (MRI) acquisition time is inherently long. Taking fewer measurements (under-sampling) can accelerate the acquisition but can lead to aliasing artifacts and loss of high-frequency information. MRI reconstruction aims to recover clinically interpretable images from the under-sampled images. Recently, deep learning-based methods have achieved state-of-the-art performances in solving MRI reconstruction problems \cite{knoll2020deep,muckley2021results,knoll2020advancing,PC-RNN,Schlemper2018,Chen2020,Duan2019}. These methods usually train neural networks to learn a mapping from under-sampled images to fully-sampled images.

\begin{figure}[t]
\centering
\includegraphics[width=0.9\textwidth]{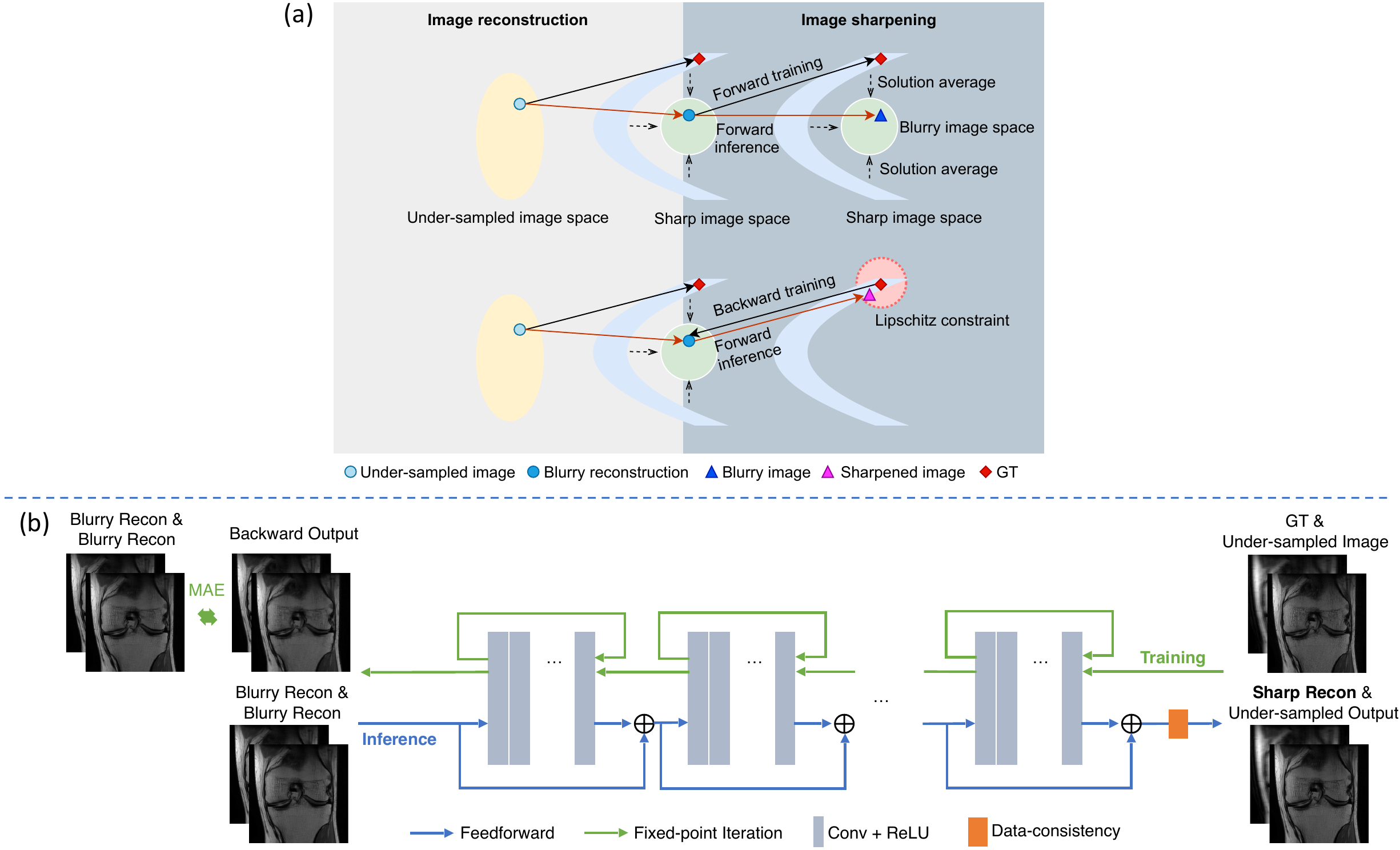}
\caption{(a) Difference between traditional methods and our method for sharpening MRI reconstruction. The MRI reconstructions are often blurry compared to the ground truth due to the solution average. The traditional forward training does not fundamentally overcome this problem and still leads to a blurry image. Our backward training utilizes a INN that learns a mapping function with Lipschitz constraints such that the predictions in the inference are close to the input in training (i.e., ground truth). (b) Architecture of InvSharpNet. During training, ground truth (GT) and the under-sampled image are passed through InvSharpNet in the inverse direction, where the fixed-point iteration algorithm is used to invert the residual blocks \cite{behrmann2019invertible}. The backward loss (Eqn.\ref{backward_loss}) is imposed between the backward output and two duplicates of the blurry reconstruction (Blurry Recon). During inference, the Blurry Recon is passed through the network in the forward direction to obtain a sharpened reconstruction (Sharp Recon).} 
\label{fig1}
\end{figure}

However, it is observed that those deep learning approaches cannot produce images as sharp as those fully-sampled images.
MRI reconstruction from under-sampled data is an ill-posed problem since it is a one-to-many problem (one under-sampled image corresponds to multiple possible fully-sampled images). Most deep learning-based methods fulfill the reconstruction task utilizing loss functions derived from metrics such as peak signal-to-noise ratio (PSNR) and structural similarity (SSIM) \cite{knoll2020advancing,muckley2021results,Jun_2021_CVPR}. To get a better PSNR, the mean-squared-error (MSE) and mean-absolute-error (MAE) losses are often used. However, as discussed in \cite{Menon2020PULSESP,Snderby2017AmortisedMI,Li2021BestBuddyGF}, training with MSE or MAE results in an image that is a pixelwise average or median of all possible solutions during network inference. This leads to an over-smoothing effect on the areas that are supposed to be rich in high-frequency details, and therefore the reconstructed images have suboptimal visual quality and might result in low diagnostic confidence. 
One way to improve the perceptual quality is adding a structural loss, which is usually achieved through maximizing SSIM \cite{Pezzotti2020,PC-RNN,GrappaNet}. However, it is still not sufficient to reach a comparable level of sharpness as the ground truth. Currently, the most popular method to improve image visual quality is adding an adversarial loss \cite{yang2021generative,malkiel2019conditional,Seitzer2018AdversarialAP}, which showed improved image visual quality compared to the methods based on only MAE and SSIM. However, it is well-known that training with adversarial loss may introduce artificial features due to the generative nature of adversarial networks. Artifacts in medical images may lead to incorrect medical diagnosis \cite{Antun2020}, which makes the adversarial loss less reliable to be implemented clinically. 

We propose a new learning framework by converting the traditional one-to-many problem to a one-to-one problem, differentiating it from previous works that propose new loss functions but still in a conventional learning fashion. We observe that given a fixed under-sampling mask, one fully-sampled image corresponds to one under-sampled image; given a fixed reconstruction network, one under-sampled image corresponds to one blurry reconstruction. 
In another observation, invertible neural networks (INN) such as \cite{behrmann2019invertible,Kingma2018GlowGF} guarantee learning a bidirectional one-to-one mapping. These observations inspire us to learn an invertible one-to-one mapping between a fully sampled image and a reconstructed image. We propose an invertible sharpening network (InvSharpNet) that can enhance the visual quality of reconstructed images through a backward training strategy from sharp fully-sampled images to blurry reconstructed images. 
To the best of our knowledge, we are the first to propose to learn an inverse mapping from ground truth to input data, which converts a one-to-many problem into a one-to-one problem and overcomes the blurry issue. Experimental results demonstrate that the backward training strategy can improve the sharpness of the images and generate fewer artifacts compared to the adversarial loss. Radiologists' evaluations also justify that our method gives a better visual quality and higher diagnostic confidence than compared methods.

\section{Methods}

\subsection{Problem Formulation}
Let $K \in \mathbb{C}^{H\times W}$ denote the fully-sampled k-space measurement with dimension of $H\times W$, and the corresponding fully-sampled image $I$ is obtained from inverse Fourier transform $I=\mathcal{F}^{-1}(K)$. 
To accelerate data acquisition, it is often to take only a portion of the k-space measurements, which is equivalent to applying an under-sampling binary mask $M$ on $K$. The corresponding under-sampled image is $Y=\mathcal{F}^{-1}(MK)$. The goal of MRI reconstruction is to find a mapping $R=\mathcal{R}(Y)$ such that $R$ is as close to $I$ as possible. The deep learning-based MRI reconstruction methods train neural networks to approximate the reconstruction mapping $\mathcal{R}$ \cite{PC-RNN,Pezzotti2020,GrappaNet}. This work aims to enhance the reconstructions $R$ by training a sharpening network $\mathcal{S}$ such that $R_{sharp}=\mathcal{S}(R)$, where $R_{sharp}$ better approximates the ground truth $I$ in terms of the visual sharpness.

\subsection{Backward Training}
Fig.\ref{fig1}(a) shows a conceptual difference between traditional forward training and our proposed backward training for sharpening MRI reconstructions. The sharp image space contains images that are visually comparable to the fully-sampled images in terms of sharpness. MRI reconstruction networks are trained to map an under-sampled image to the ground truth, which is located in the sharp image space. However, training with a pixelwise loss (MAE or MSE) maps the under-sampled image to the blurry image space due to the solution average problem. 

The traditional method to improve the blurry reconstruction is training a refinement network in the forward direction that maps the reconstruction to the ground truth \cite{Seitzer2018AdversarialAP,Quan2018CompressedSM}. One way to achieve this is training the refinement network with a weighted sum of MAE and structural loss based on Multiscale SSIM (MSSIM) \cite{Wang} between the network output and ground truth \cite{Pezzotti2020}: 
\begin{equation}
\mathcal{L}_{forward}=(1-\alpha)||\mathcal{S}(R)-I||_{1}+\alpha(1-\text{MSSIM}(\mathcal{S}(R),I))
\label{forloss}
\end{equation}
However, this forward training cannot overcome the solution average issue and the output image still resides in the blurry image space (Fig.\ref{fig1}(a)). Therefore, the refinement with forward training does not improve the visual quality. 

We propose an InvSharpNet that adopts backward training to learn an image blurring transform. Specifically, the ground truth image is fed into the output side of InvSharpNet and then inversely passed through the network to obtain an image at the input side, leveraging the network's invertibility. The MAE between this backward output and the blurry reconstruction is minimized:
\begin{equation}
\mathcal{L}_{backward}=||\mathcal{S}^{-1}(I, Y)-R||_1
\label{backward_loss}
\end{equation}
The under-sampled image $Y$ is also provided as the input to provide information about under-sampling mask (see Section \ref{section_architecture}). As shown in Fig.\ref{fig1}(a), the backward training learns a blurring transform from the ground truth to the blurry reconstruction. At inference, InvSharpNet is used in the forward direction such that the blurry reconstruction is regarded as the network's input. InvSharpNet is designed based on iResNet \cite{behrmann2019invertible} which imposes Lipschitz constraint on the network layers. The Lipschitz constraint guarantees that a small difference in network's input will not result in a large difference in network's output, meaning that the output in inference is close to the input in training. Therefore, InvSharpNet can obtain a sharpened image close to the ground truth in the sharp image space. 

\subsection{Network Architecture}
\label{section_architecture}
Our InvSharpNet, shown in Fig.\ref{fig1}(b), contains 12 convolution blocks, and each has 5 convolution layers with 128 channels in each intermediate layer.  

\textbf{Lipschitz constraint} The Lipschitz constant of each convolution layer should be less than 1 to achieve full invertibility, and the inverse is computed via a fixed-point iteration algorithm \cite{behrmann2019invertible}. Given the output of each convolution block, the input is computed by iteratively looping through the block. A small Lipschitz constant, e.g. $c=0.5$, requires fewer iterations to guarantee full invertibility but significantly inhibits the network's learning capability \cite{behrmann2019invertible}. A large Lipschitz constant, e.g. $c=0.9$, retains greater learning capability but requires more iterations to compute the inverse. One more iteration is equivalent to a 1-fold increase in network size. Therefore, a large $c$ could significantly increase the memory usage and training time. We choose $c=0.7$ and 2 iterations to balance the trade-off between learning capability and training time (see Section \ref{section_experimental_results}). 

\textbf{Conditioning on under-sampling mask} In order to learn a mapping from the ground truth to a specific reconstruction that depends on the under-sampling mask, we let InvSharpNet condition on the corresponding mask. This can be achieved by concatenating the ground truth with the under-sampled image as the input during training, since the under-sampled image is generated with the under-sampling mask $M$ and thus contains the mask information. 
As INN requires the same dimensionality on input and output, the network outputs two images, and we train both of them to approach the blurry reconstruction using Eqn.\ref{backward_loss}. During inference, two duplicates of the reconstruction are input to the network in the forward direction. The output corresponding to the ground truth during training is taken as the sharpened reconstruction. 


\textbf{Data fidelity} A DC layer is appended at the network's output in the forward (inference) direction to make the prediction consistent with measurements \cite{Duan2019}.

\section{Experiments}
\label{sec:experiments}
We demonstrate the sharpening performance of InvSharpNet based on reconstructions from two models: PC-RNN \cite{PC-RNN}, which ranked among the best-performed methods in 2019 fastMRI challenge \cite{knoll2020advancing}, and U-Net \cite{UNet}. We name these two reconstruction models as recon model 1 and 2, respectively.

We first implemented a refinement network that only uses the forward training (Eqn.\ref{forloss}) with $\alpha$ set to 0.84 as suggested in \cite{Pezzotti2020}. We also implemented the conditional Wasserstein GAN (cWGAN) \cite{malkiel2019conditional,Seitzer2018AdversarialAP,Quan2018CompressedSM} by training with a weighted sum of forward and adversarial loss:
\begin{equation}
    \mathcal{L}_{cWGAN}=\mathcal{L}_{forward}+\gamma \mathcal{L}_{adv}
\end{equation}
where $\gamma$ controls the trade-off between evaluation metrics and visual quality. $\gamma=0.01$ and $\gamma=0.02$ were tested to observe the difference. After that, we implemented two versions of InvSharpNet: one uses the backward training proposed in Eqn.\ref{backward_loss} and the other one uses a bidirectional training 
\begin{equation}
\mathcal{L}_{bi}=\mathcal{L}_{forward}+\beta\mathcal{L}_{backward}
\label{biloss}
\end{equation}
with $\beta$ set to 2. The bidirectional training can be regarded as a compromise between evaluation metrics and visual quality. We used the same network architecture, illustrated in Fig.\ref{fig1}(b), for all methods. Trainings were performed with 50k iterations, batch size of 8 and a learning rate of 0.001 with Adam optimizer. Experiments were implemented in PyTorch on NVIDIA Tesla V100 GPUs. 

\subsection{Datasets}
\textbf{FastMRI knee dataset } \cite{fastmri} includes 34,742 2D images from 973 subjects for training and 7,135 images from 199 subjects for evaluation, including modalities of proton-density weighted images with and without fat suppression. In this work, we focus on 4X acceleration of multi-coil data (same for fastMRI brain).

\textbf{FastMRI brain dataset} \cite{fastmri} includes 67,558 images from 4,469 subjects for training and 20,684 images from 1,378 subjects for evaluation, including modalities of T1-weighted, T1-weighted pre-contrast, T1-weighted post-contrast, T2-weighted and T2 Fluid-attenuated inversion recovery images. 

\textbf{In-house brain dataset} contains 584 images for evaluation, including five different gradient echo (GRE) contrasts. The data were collected with Institutional Reviews Board (IRB). 
We directly evaluate the networks trained on the fastMRI brain dataset.

\subsection{Results}
\label{section_experimental_results}
\begin{figure}[t]
\centering
\includegraphics[width=0.9\textwidth]{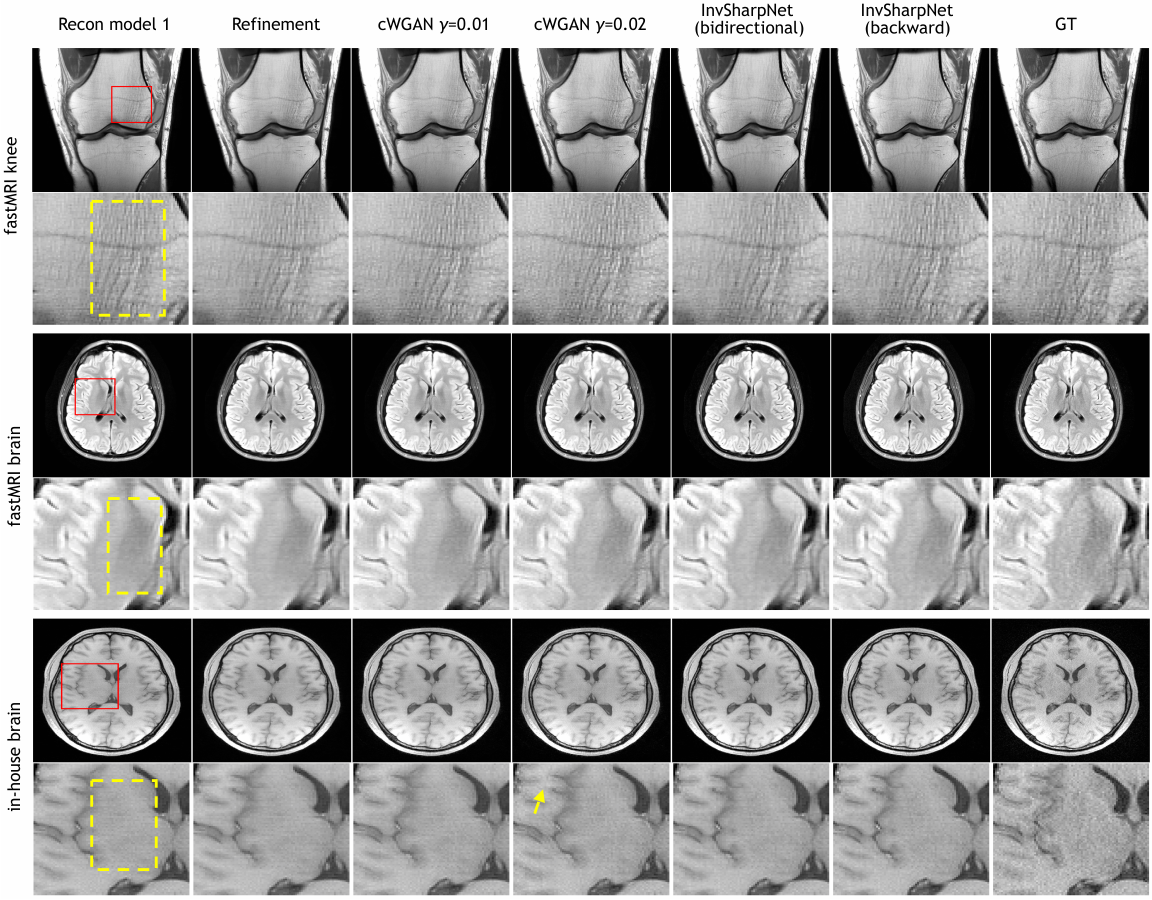}
\caption{Qualitative comparisons. The red boxes indicate the zoom-in areas and the yellow dashed boxes show regions where sharpness difference can be observed. The yellow arrow is an example of artifacts introduced by cWGAN. GT = Ground Truth.} 
\label{fig2}
\end{figure}

\textbf{Qualitative Measure} Fig.\ref{fig2} illustrates examples of the sharpening results given by the compared methods on both public fastMRI and in-house datasets. The first column shows that the reconstructions given by recon model 1 look blurry compared to the ground truth images in the last column. As shown in the second column, training a refinement network in the forward direction produces almost identical images to the original reconstructions and fails to improve the visual quality. As well-documented in previous literature, training with the adversarial loss (cWGAN $\gamma=0.01$) can improve the image sharpness, and using a stronger weight ($\gamma=0.02$) can achieve further improvement. Using InvSharpNet with bidirectional training achieved a similar level of sharpness enhancement as using cWGAN $\gamma=0.01$. Finally, using InvSharpNet with backward training achieved even better sharpness than the bidirectional training. To better understand the visual quality improvement, we provide visualizations of k-space and image profile in Appendix Fig.A1. We also show a pathology case in Appendix Fig.A2. 

The main advantage of using InvSharpNet instead of the cWGAN is the lower risk of introducing artifacts. As shown in Fig.\ref{fig2} and Appendix Fig.A3, using generative models like cWGAN introduces artifacts (indicated by the yellow arrows) that were neither originally contained in the reconstructions nor contained in the ground truth, increasing the risk of incorrect medical diagnosis. Additionally, our results were evaluated from a clinical perspective using radiologists' ratings, which was considered as a key evaluation criterion in fastMRI competitions \cite{knoll2020advancing,muckley2021results}. 2 radiologists examined the image quality of 6 selected cases (156 2D images) from the fastMRI knee and brain datasets in a blind fashion. The fully sampled images are given as rating references. The radiologists rated each image by artifacts, sharpness, contrast-to-noise ratio, diagnostic confidence and an overall score. The results are reported in Table \ref{radiologist}, which shows that InvSharpNet with backward training outperforms other methods from all aspects. Although cWGAN models can also give higher sharpness scores than original reconstructions, they introduce artifacts that result in lower artifacts scores. 

\begin{table}[t]
    \tiny
    \centering
    \caption{Evaluation from 2 radiologists in a blind fashion over 3 knee and 3 brain cases from the fastMRI dataset. Ratings follow a 0-5 point scale, where 5 is the best and 0 is the worst. Results are shown in mean$\pm$SD.}
    \begin{tabular}{ c | c  c  c  c  c }
         Method & Artifacts & Sharpness & Contrast-to-noise & Diagnostic confidence & Overall \\
         \hline
         Recon model 1 & \textbf{4.00$\pm$0.95} & 4.00$\pm$0.71 & 4.25$\pm$0.40 & 3.75$\pm$1.06 & 3.67$\pm$0.98\\
         Refinement & 3.92$\pm$0.87 & 3.88$\pm$0.64 & 4.21$\pm$0.40 & 3.71$\pm$1.01 & 3.67$\pm$0.98\\
         cWGAN $\gamma$=0.01 & 3.75$\pm$0.89 & 4.17$\pm$0.58 & 4.33$\pm$0.39 & 3.75$\pm$1.01 & 3.71$\pm$0.96\\
         cWGAN $\gamma$=0.02 & 3.58$\pm$0.95 & 4.33$\pm$0.44 & 4.29$\pm$0.45 & 3.79$\pm$1.08 & 3.79$\pm$0.96\\
         InvSharpNet (bidirectional) & 3.96$\pm$0.92 & 4.21$\pm$0.62 & 4.25$\pm$0.26 & \textbf{3.88$\pm$1.13} & 3.71$\pm$0.96\\
         InvSharpNet (backward) & \textbf{4.00$\pm$0.88} & \textbf{4.67$\pm$0.54} & \textbf{4.38$\pm$0.43} & \textbf{3.88$\pm$1.13} & \textbf{3.88$\pm$1.07}\\
         \hline
    \end{tabular}
    \label{radiologist}
\end{table}


\textbf{Quantitative Measure} Table \ref{metric} reports the commonly used evaluation metrics PSNR and SSIM. However, PSNR and SSIM are often degraded as sharpness improves due to the well-known perception-distortion tradeoff \cite{blau2018perception}, making those metrics ineffective in evaluating our improvement. We observed that the image sharpness is directly related to the contrast term in SSIM \cite{Wang2004ImageQA}, so we also report it in Table \ref{metric}.
The refinement network gives the best PSNR and SSIM among all methods because it just focuses on minimizing the pixelwise and structural loss between the network's output and the ground truth. However, the contrast metric ranks the lowest for all datasets, indicating that high PSNR and SSIM scores do not necessarily correlate to good visual quality. Radiologists' ratings in Table \ref{radiologist} confirm this point: the refinement network provides images with the lowest sharpness and diagnostic confidence. On the other hand, although our InvSharpNet gives lower PSNR and SSIM than the refinement network, it provides images with better contrast and higher radiologists' ratings. Training InvSharpNet with a bidirectional strategy is a way to balance the trade-off between commonly used evaluation metrics and visual quality. A similar trade-off can also be observed when different values of $\gamma$ were used for cWGAN.

\begin{table}[t]
  \tiny
  \centering
  \caption{Quantitative comparisons using PSNR, SSIM and Contrast. Best scores are shown in bold. Underline indicates the second best contrast.}
  \begin{tabular}{>{\centering\arraybackslash}p{1.95cm} | >{\centering\arraybackslash}p{0.9cm} >{\centering\arraybackslash}p{1.05cm} >{\centering\arraybackslash}p{1.1cm} | >{\centering\arraybackslash}p{0.94cm} >{\centering\arraybackslash}p{1.05cm} >{\centering\arraybackslash}p{1.1cm} | >{\centering\arraybackslash}p{0.92cm} >{\centering\arraybackslash}p{1.05cm} >{\centering\arraybackslash}p{1.1cm}}
     & \multicolumn{3}{c|}{fastMRI knee} & \multicolumn{3}{c|}{fastMRI brain} & 
     \multicolumn{3}{c}{in-house brain} \\
    \hline
    Method & PSNR & SSIM & Contrast & PSNR & SSIM & Contrast & PSNR & SSIM & Contrast \\ 
    \hline
    Recon model 1 & 38.9$\pm$2.8 & .921$\pm$.043 & .987$\pm$.008 & 40.2$\pm$2.2 & .958$\pm$.015 & .996$\pm$.002 & \textbf{32.5$\pm$6.0} & .825$\pm$.175 & .974$\pm$.034 \\
    Refinement & \textbf{39.1$\pm$2.7} & \textbf{.929$\pm$.039} & .984$\pm$.011 & \textbf{40.6$\pm$2.3} & \textbf{.968$\pm$.014} & .994$\pm$.003 & 32.3$\pm$5.6 & \textbf{.850$\pm$.165} & .964$\pm$.057\\
    cWGAN $\gamma$=0.01 & 38.9$\pm$2.7 & .927$\pm$.039 & .990$\pm$.006 & 40.3$\pm$2.3 & .967$\pm$.014 & .996$\pm$.003 & 31.9$\pm$5.5 & .848$\pm$.161 & .972$\pm$.043 \\
    cWGAN $\gamma$=0.02 & 38.3$\pm$2.6 & .920$\pm$.041 & \textbf{.993$\pm$.004} & 39.6$\pm$2.2 & .963$\pm$.015 & \textbf{.997$\pm$.002} & 31.5$\pm$5.2 & .840$\pm$.161 & \underline{.980$\pm$.029} \\
    InvSharpNet(bi)& 38.8$\pm$2.8 & .922$\pm$.042 & .989$\pm$.007 & 40.2$\pm$2.2 & .961$\pm$.015 & \underline{.997$\pm$.002} & 32.5$\pm$5.9 & .835$\pm$.174 & .977$\pm$.030\\
    InvSharpNet(back)& 38.5$\pm$2.7 & .920$\pm$.043 & \underline{.992$\pm$.005} & 39.4$\pm$2.1 & .950$\pm$.017 & .996$\pm$.002 & 32.4$\pm$5.7 & .841$\pm$.170 & \textbf{.981$\pm$.025} \\
    \hline
  \end{tabular}
  \label{metric}
\end{table}

We further demonstrate the effectiveness of our method using LPIPS \cite{zhang2018unreasonable}, which measures images' high-level similarity and correlates well with human perceptual judgment (Table \ref{lpips}). InvSharpNet achieves better LPIPS on all three datasets compared to the reconstruction model and the refinement network, consistent with radiologists' ratings (Table \ref{radiologist}) and the contrast metric (Table \ref{metric}). 

\begin{table}[b]
  \tiny
  \centering
  \caption{Quantitative comparisons using LPIPS (lower means more similar to GT).}
  \begin{tabular}{>{\centering\arraybackslash}p{2.5cm} |  >{\centering\arraybackslash}p{2.0cm} |  >{\centering\arraybackslash}p{2.0cm} |  >{\centering\arraybackslash}p{2.0cm}}
    Method & fastMRI knee & fastMRI brain & in-house brain \\
    \hline
    Recon model 1 & 0.078$\pm$0.032 & 0.035$\pm$0.015 & 0.067$\pm$0.074\\
    Refinement & 0.089$\pm$0.037 & 0.041$\pm$0.018 & 0.078$\pm$0.091\\
    InvSharpNet(bi) & 0.070$\pm$0.029 & \textbf{0.028$\pm$0.013} & 0.058$\pm$0.069\\
    InvSharpNet(back) & \textbf{0.068$\pm$0.025} & \textbf{0.028$\pm$0.012} & \textbf{0.055$\pm$0.064}\\
    \hline
  \end{tabular}
  \label{lpips}
\end{table}

\begin{figure}[t]
\centering
\includegraphics[width=0.9\textwidth]{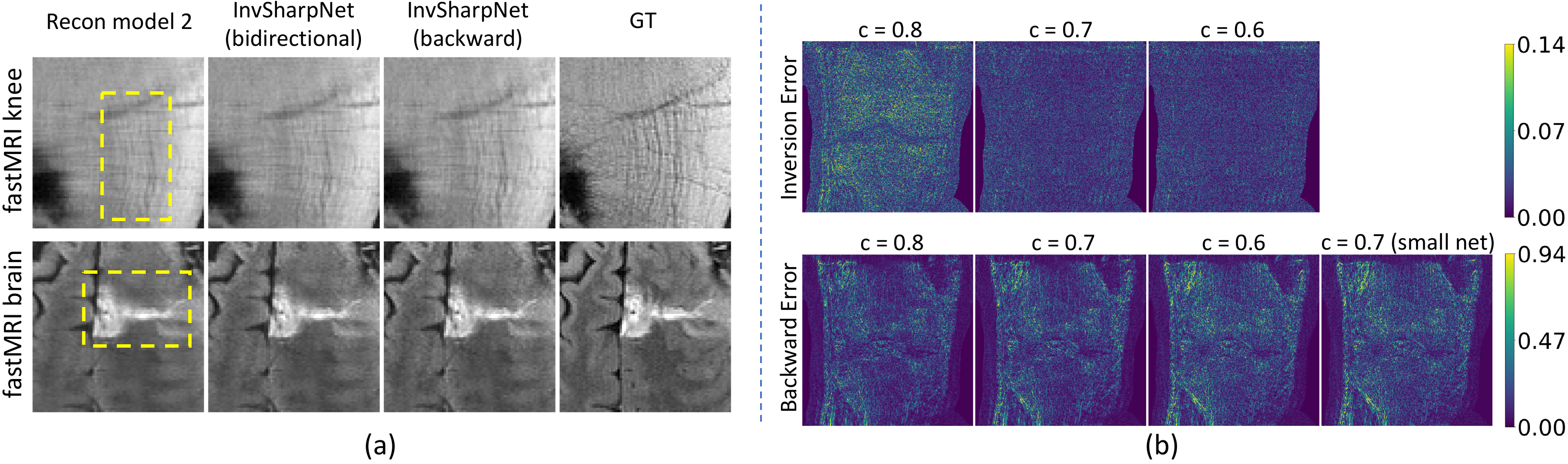}
\caption{(a) Generalizability. Sharpening results based on reconstructions from recon model 2. (b) Ablation studies on Lipschitz constants $c$ and network sizes. } 
\label{fig3}
\end{figure}

\textbf{Generalizability} 
We also obtained results of sharpening the reconstructions based on recon model 2 in Fig.\ref{fig3}(a). Similar sharpness enhancement can be observed for reconstructions from recon model 2, which shows that InvSharpNet is a generalizable method. Evaluation scores in Appendix Table A1 further validate that InvSharpNet can improve the contrast of recon model 2 reconstructions. 

\textbf{Lipschitz Constant} We compare three Lipschitz constants $c=0.8,0.7,0.6$ by fixing the number of fixed-point iterations to 2 due to limited computation resources. Fig.\ref{fig3}(b)(top) shows the inversion error defined as $E_{inv} = |I - \mathcal{S}(\mathcal{S}^{-1}(I))|$. A large inversion error means that a backward pass through the network is a poor approximation of the inverse of the forward pass, which could result in erroneous mapping learned from the backward training. $c=0.8$ is a suboptimal choice as it gives an inversion error map in which the knee structure can clearly be observed, and $c=0.7$ and $0.6$ result in much smaller inversion errors. We then investigate how Lipschitz constant affects network's learning capability by showing the backward error as defined in Eqn.\ref{backward_loss} for one image example (Fig.\ref{fig3}(b)(bottom)). The backward error progressively increases as $c$ is reduced from 0.8 to 0.6, implying a degradation of learning capability. To balance the inversion error and network's learning capability, we chose $c=0.7$.

\textbf{Network Size} A large network size is required to compensate for the Lipschitz constraint that undermines networks' learning capability. InvSharpNet contains roughly 5.4M parameters, and we justify this choice by comparing it with a smaller network containing 0.7M parameters. We show that when $c=0.7$, using the small network results in a larger backward error (Fig.\ref{fig3}(b)(bottom)), indicating that the mapping is not learned as well as the larger model. 

\section{Conclusion}
We propose a novel InvSharpNet that learns a blurring transform from the fully-sampled MRI images to the blurry reconstructions, which is inverted during inference to enhance the blurry input. Results show that InvSharpNet can improve image quality given by the existing reconstruction methods, providing higher diagnostic confidence for clinicians. The method can be extended to image denoising and super-resolution. 

%
%
%
\bibliographystyle{splncs04}
\bibliography{reference}

\begin{thebibliography}{10}
\providecommand{\url}[1]{\texttt{#1}}
\providecommand{\urlprefix}{URL }
\providecommand{\doi}[1]{https://doi.org/#1}

\bibitem{Antun2020}
Antun, V., Renna, F., Poon, C., Adcock, B., Hansen, A.C.: On instabilities of
  deep learning in image reconstruction and the potential costs of ai.
  Proceedings of the National Academy of Sciences  \textbf{117}(48),
  30088--30095 (2020)

\bibitem{behrmann2019invertible}
Behrmann, J., Grathwohl, W., Chen, R.T., Duvenaud, D., Jacobsen, J.H.:
  Invertible residual networks. In: International Conference on Machine
  Learning. pp. 573--582. PMLR (2019)

\bibitem{blau2018perception}
Blau, Y., Michaeli, T.: The perception-distortion tradeoff. In: Proceedings of
  the IEEE conference on computer vision and pattern recognition. pp.
  6228--6237 (2018)

\bibitem{Chen2020}
Chen, E.Z., Chen, T., Sun, S.: Mri image reconstruction via learning
  optimization using neural odes. In: International Conference on Medical Image
  Computing and Computer-Assisted Intervention. pp. 83--93. Springer (2020)

\bibitem{Duan2019}
Duan, J., Schlemper, J., Qin, C., Ouyang, C., Bai, W., Biffi, C., Bello, G.,
  Statton, B., O’regan, D.P., Rueckert, D.: Vs-net: Variable splitting
  network for accelerated parallel mri reconstruction. In: International
  Conference on Medical Image Computing and Computer-Assisted Intervention. pp.
  713--722. Springer (2019)

\bibitem{Jun_2021_CVPR}
Jun, Y., Shin, H., Eo, T., Hwang, D.: Joint deep model-based mr image and coil
  sensitivity reconstruction network (joint-icnet) for fast mri. In:
  Proceedings of the IEEE/CVF Conference on Computer Vision and Pattern
  Recognition. pp. 5270--5279 (2021)

\bibitem{Kingma2018GlowGF}
Kingma, D.P., Dhariwal, P.: Glow: Generative flow with invertible 1x1
  convolutions. Advances in neural information processing systems  \textbf{31}
  (2018)

\bibitem{knoll2020deep}
Knoll, F., Hammernik, K., Zhang, C., Moeller, S., Pock, T., Sodickson, D.K.,
  Akcakaya, M.: Deep-learning methods for parallel magnetic resonance imaging
  reconstruction: A survey of the current approaches, trends, and issues. IEEE
  signal processing magazine  \textbf{37}(1),  128--140 (2020)

\bibitem{knoll2020advancing}
Knoll, F., Murrell, T., Sriram, A., Yakubova, N., Zbontar, J., Rabbat, M.,
  Defazio, A., Muckley, M.J., Sodickson, D.K., Zitnick, C.L., et~al.: Advancing
  machine learning for mr image reconstruction with an open competition:
  Overview of the 2019 fastmri challenge. Magnetic resonance in medicine
  \textbf{84}(6),  3054--3070 (2020)

\bibitem{fastmri}
Knoll, F., Zbontar, J., Sriram, A., Muckley, M.J., Bruno, M., Defazio, A.,
  Parente, M., Geras, K.J., Katsnelson, J., Chandarana, H., et~al.: fastmri: A
  publicly available raw k-space and dicom dataset of knee images for
  accelerated mr image reconstruction using machine learning. Radiology:
  Artificial Intelligence  \textbf{2}(1),  e190007 (2020)

\bibitem{Li2021BestBuddyGF}
Li, W., Zhou, K., Qi, L., Lu, L., Jiang, N., Lu, J., Jia, J.: Best-buddy gans
  for highly detailed image super-resolution. arXiv preprint arXiv:2103.15295
  (2021)

\bibitem{malkiel2019conditional}
Malkiel, I., Ahn, S., Taviani, V., Menini, A., Wolf, L., Hardy, C.J.:
  Conditional wgans with adaptive gradient balancing for sparse mri
  reconstruction. arXiv preprint arXiv:1905.00985  (2019)

\bibitem{Menon2020PULSESP}
Menon, S., Damian, A., Hu, S., Ravi, N., Rudin, C.: Pulse: Self-supervised
  photo upsampling via latent space exploration of generative models. In:
  Proceedings of the ieee/cvf conference on computer vision and pattern
  recognition. pp. 2437--2445 (2020)

\bibitem{muckley2021results}
Muckley, M.J., Riemenschneider, B., Radmanesh, A., Kim, S., Jeong, G., Ko, J.,
  Jun, Y., Shin, H., Hwang, D., Mostapha, M., et~al.: Results of the 2020
  fastmri challenge for machine learning mr image reconstruction. IEEE
  transactions on medical imaging  \textbf{40}(9),  2306--2317 (2021)

\bibitem{Pezzotti2020}
Pezzotti, N., Yousefi, S., Elmahdy, M.S., Van~Gemert, J.H.F., Schuelke, C.,
  Doneva, M., Nielsen, T., Kastryulin, S., Lelieveldt, B.P., Van~Osch, M.J.,
  et~al.: An adaptive intelligence algorithm for undersampled knee mri
  reconstruction. IEEE Access  \textbf{8},  204825--204838 (2020)

\bibitem{Quan2018CompressedSM}
Quan, T.M., Nguyen-Duc, T., Jeong, W.K.: Compressed sensing mri reconstruction
  using a generative adversarial network with a cyclic loss. IEEE transactions
  on medical imaging  \textbf{37}(6),  1488--1497 (2018)

\bibitem{UNet}
Ronneberger, O., Fischer, P., Brox, T.: U-net: Convolutional networks for
  biomedical image segmentation. In: International Conference on Medical image
  computing and computer-assisted intervention. pp. 234--241. Springer (2015)

\bibitem{Schlemper2018}
Schlemper, J., Caballero, J., Hajnal, J.V., Price, A.N., Rueckert, D.: A deep
  cascade of convolutional neural networks for dynamic mr image reconstruction.
  IEEE transactions on Medical Imaging  \textbf{37}(2),  491--503 (2017)

\bibitem{Seitzer2018AdversarialAP}
Seitzer, M., Yang, G., Schlemper, J., Oktay, O., W{\"u}rfl, T., Christlein, V.,
  Wong, T., Mohiaddin, R., Firmin, D., Keegan, J., et~al.: Adversarial and
  perceptual refinement for compressed sensing mri reconstruction. In:
  International conference on medical image computing and computer-assisted
  intervention. pp. 232--240. Springer (2018)

\bibitem{Snderby2017AmortisedMI}
S{\o}nderby, C.K., Caballero, J., Theis, L., Shi, W., Husz{\'a}r, F.: Amortised
  map inference for image super-resolution. arXiv preprint arXiv:1610.04490
  (2016)

\bibitem{GrappaNet}
Sriram, A., Zbontar, J., Murrell, T., Zitnick, C.L., Defazio, A., Sodickson,
  D.K.: Grappanet: Combining parallel imaging with deep learning for multi-coil
  mri reconstruction. In: Proceedings of the IEEE/CVF Conference on Computer
  Vision and Pattern Recognition. pp. 14315--14322 (2020)

\bibitem{PC-RNN}
Wang, P., Chen, E.Z., Chen, T., Patel, V.M., Sun, S.: Pyramid convolutional rnn
  for mri reconstruction. arXiv preprint arXiv:1912.00543  (2019)

\bibitem{Wang2004ImageQA}
Wang, Z., Bovik, A.C., Sheikh, H.R., Simoncelli, E.P.: Image quality
  assessment: from error visibility to structural similarity. IEEE transactions
  on image processing  \textbf{13}(4),  600--612 (2004)

\bibitem{Wang}
Wang, Z., Simoncelli, E.P., Bovik, A.C.: Multiscale structural similarity for
  image quality assessment. In: The Thrity-Seventh Asilomar Conference on
  Signals, Systems \& Computers, 2003. vol.~2, pp. 1398--1402. Ieee (2003)

\bibitem{yang2021generative}
Yang, G., Lv, J., Chen, Y., Huang, J., Zhu, J.: Generative adversarial networks
  (gan) powered fast magnetic resonance imaging--mini review, comparison and
  perspectives. arXiv preprint arXiv:2105.01800  (2021)

\bibitem{zhang2018unreasonable}
Zhang, R., Isola, P., Efros, A.A., Shechtman, E., Wang, O.: The unreasonable
  effectiveness of deep features as a perceptual metric. In: Proceedings of the
  IEEE conference on computer vision and pattern recognition. pp. 586--595
  (2018)

\end{thebibliography}

\centering{\Large\textbf{Appendix}}
\renewcommand\thefigure{A\arabic{figure}} 
\renewcommand\thetable{A\arabic{table}} 

\setcounter{figure}{0} 
\setcounter{table}{0} 

\begin{figure}
\centering
\includegraphics[width=\textwidth]{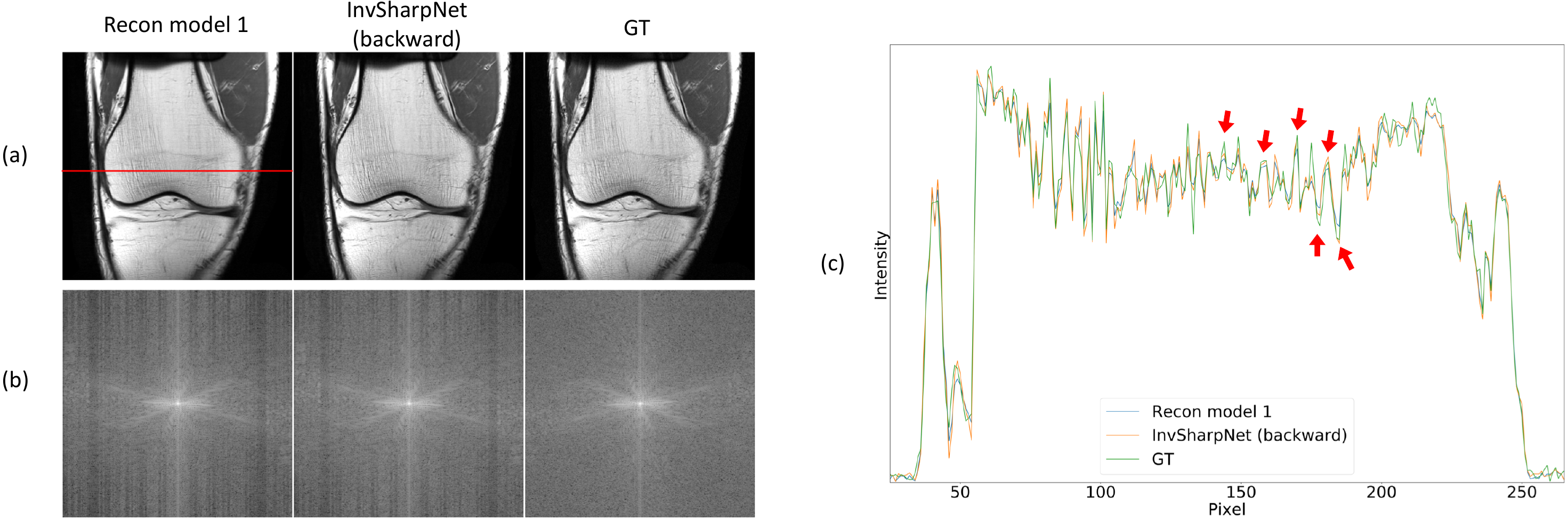}
\caption{(a) Image domain comparison. (b) k-space comparison. The fully-sampled k-space is smooth, whereas the k-space reconstructed by recon model 1 has lower energy at the unmeasured positions (mostly high-frequency components). Using InvSharpNet renders a sharper image with higher energy for high-frequency components. (c) Image profile at the red horizontal line in (a). InvSharpNet pulls the pixel values closer to the ground truth at some over-smoothed regions (the red arrows indicate a few examples).} 
\label{figA1}
\end{figure}

\begin{figure}
\centering
\includegraphics[width=\textwidth]{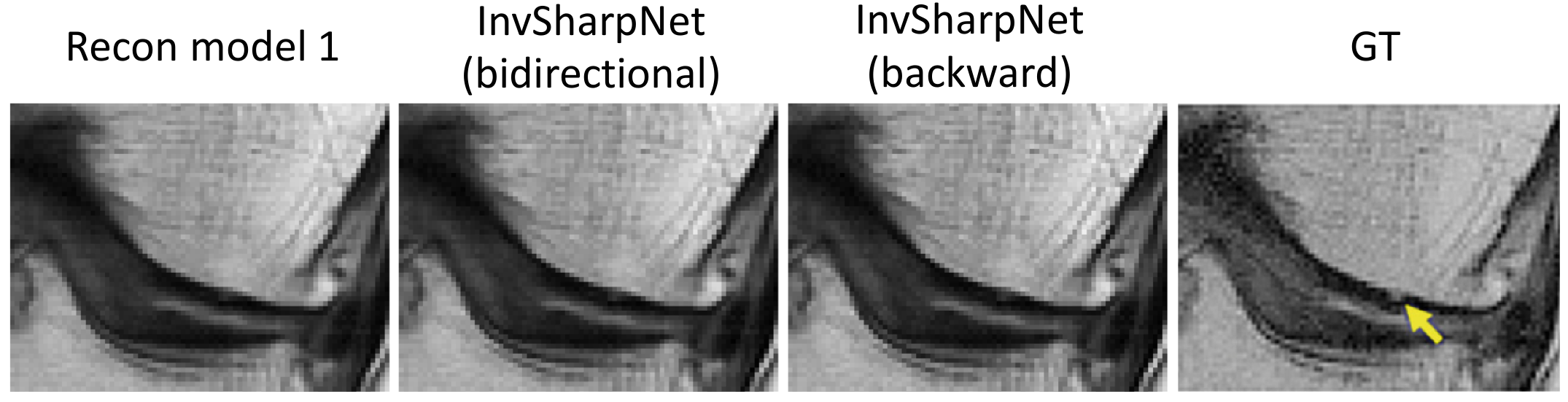}
\caption{Comparison of different methods on recovering a pathology. The ground truth (GT) is a screenshot from (Johnson et al., 2021), in which the yellow arrow indicates the pathology. Using InvSharpNet with backward or bidirectional training can recover better image contrast at the pathology than the original reconstruction, providing higher diagnostic confidence.} 
\label{figA2}
\end{figure}

\begin{figure}
\centering
\includegraphics[width=\textwidth]{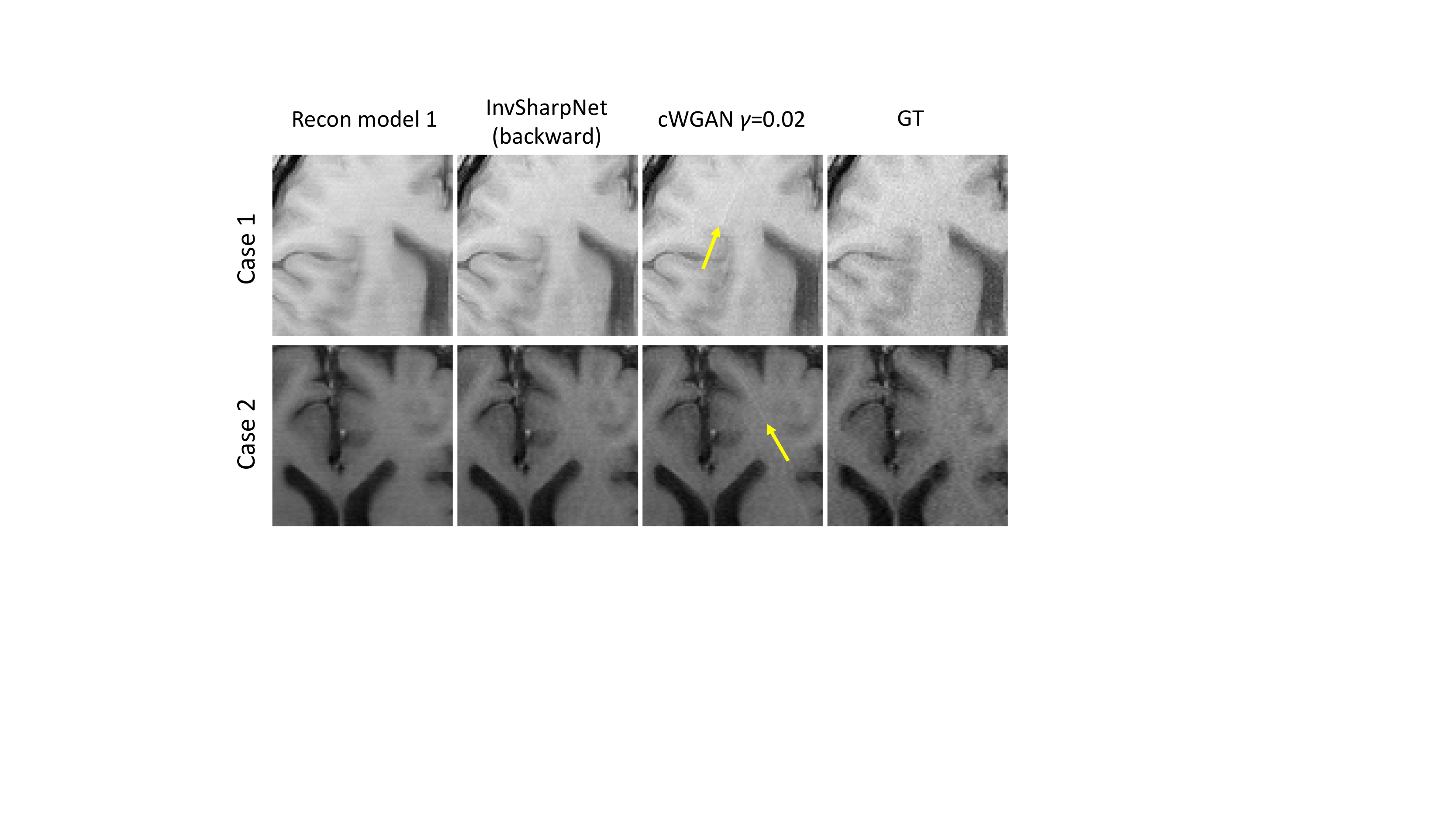}
\caption{Examples of artifacts introduced by cWGAN but not by InvSharpNet.} 
\label{figA3}
\end{figure}

\begin{table}
  \tiny
  \centering
  \caption{Quantitative evaluation of sharpening results based on reconstructions from recon model 2.}
  \begin{tabular}{>{\centering\arraybackslash}p{1.95cm} | >{\centering\arraybackslash}p{0.9cm} >{\centering\arraybackslash}p{1.05cm} >{\centering\arraybackslash}p{1.1cm} | >{\centering\arraybackslash}p{0.9cm} >{\centering\arraybackslash}p{1.05cm} >{\centering\arraybackslash}p{1.1cm} | >{\centering\arraybackslash}p{0.9cm} >{\centering\arraybackslash}p{1.05cm} >{\centering\arraybackslash}p{1.1cm}}
     & \multicolumn{3}{c|}{fastMRI knee} & \multicolumn{3}{c|}{fastMRI brain} & 
     \multicolumn{3}{c}{in-house brain} \\
    \hline
    Method & PSNR & SSIM & Contrast & PSNR & SSIM & Contrast & PSNR & SSIM & Contrast \\ 
    \hline
    Recon model 2 & 35.5$\pm$1.9 & .886$\pm$.041 & .977$\pm$.006 & \textbf{36.3$\pm$2.2} & \textbf{.932$\pm$.019} & .990$\pm$.003 & \textbf{30.8$\pm$5.5} & \textbf{.811$\pm$.173} & .972$\pm$.031 \\
    InvSharpNet(bi) & \textbf{35.6$\pm$1.9} & \textbf{.888$\pm$.040} & \underline{.979$\pm$.006} & 36.1$\pm$2.2 & .928$\pm$.020 & \textbf{.992$\pm$.003} & 30.6$\pm$5.3 & .811$\pm$.168 & \underline{.977$\pm$.025} \\
    InvSharpNet(back) & 35.3$\pm$2.0 & .879$\pm$.046 & \textbf{.984$\pm$.005} & 35.6$\pm$2.1 & .911$\pm$.024 & \underline{.991$\pm$.003} & 30.2$\pm$5.1 & .804$\pm$.157 & \textbf{.980$\pm$.018} \\
    \hline
  \end{tabular}
  \label{tabA2}
\end{table}

\end{document}